\documentstyle[twocolumn,epsf]{mn}

\title{ On the self-consistent model of the axisymmetric
radio pulsar magnetosphere}

\author[V.S.~Beskin and L.M.~Malyshkin]
       {V.S.~Beskin$^1$ and L.M.Malyshkin$^2$ \\
	   $^1$ P.N.Lebedev Physical Institute, Leninsky prosp., 53,
Moscow, 117924, Russia\\
	   $^2$ Moscow Institute of Physics and Technology,
Dolgoprudny, Moscow region, 141700, Russia}
\date{Accepted 1997  .
     Received  1997 ;
     in original form 1997 }

\def\be{\begin{equation}}
\def\ee{\end{equation}}
\def\beq{\begin{eqnarray}}
\def\eeq{\end{eqnarray}}

\pagerange{\pageref{firstpage}--\pageref{lastpage}}

\pubyear{1997}

\begin{document}

\maketitle

\label{firstpage}

\begin{abstract}
We consider a model of axisymmetric neutron star magnetosphere. In our
approach, the current density in the region of open field lines is constant
and the returning current flows in a narrow layer along the separatrix. In
this case, the stream equation describing the magnetic field structure is linear
both in the open and closed regions, the main problem lying in matching the
solutions along the separatrix (Okamoto 1974; Lyubarskii 1990).
We demonstrate that it is the stability
condition on the separatrix that allows us to obtain a unique solution of the
problem. In particular, the zero point of magnetic field is shown to locate
near the light cylinder. Moreover, the hypothesis of the
existence of the nonlinear Ohm's Law (Beskin, Gurevich \& Istomin 1983)
 connecting the potential drop in the pair creation region and the
longitudinal electric current flowing in the magnetosphere is confirmed.
\end{abstract}


\section{Introduction}

A few years after the discovery of radio pulsars  (Hewish et al 1968)
the basic properties of the neutron star magnetosphere were actually
clarified.  Firstly, the importance of one-photon electron-positron pair
creation in a superstrong magnetic field was shown (Sturrock 1971; Ruderman
\& Sutherland 1975). It means that the magnetosphere of a neutron star
must be filled with
a plasma screening the longitudinal electric field (Goldreich \& Julian 1969),
the charge density $\rho_{e}$ being nonzero. This screening results in the
corotation of a plasma with the star.  Clearly, such a corotation is
impossible outside the light cylinder $R_{L}=c/\Omega$, where $\Omega$ is the
angular velocity of a star. Hence, in the magnetosphere there form two
essentially different groups of field lines:  closed, i.e. those returning to
the star surface, and open, i.e. those crossing the light cylinder and going
to infinity.  As a result, along open field lines the plasma may leave
the neutron star and escape from the magnetosphere. The charge $\rho_{e}$
going together with plasma produces a longitudinal electric current.  It was
also shown that it is the ponderomotive action of the electric current flowing
along open field lines and closing on the star surface that diminishes the
angular velocity of a pulsar.

At the same time, the equation describing the structure of an axisymmetric
stationary magnetosphere in the force-free approximation was obtained by
several authors (Mestel 1973; Scharlemann \& Wagoner 1973; Michel 1973ab;
Okamoto 1974).  This second-order elliptical equation on the magnetic flux
function has the Grad-Shafranov form i.e. contains two `integrals of
motion', a longitudinal electric current and an angular velocity of magnetic
surfaces, which are constant along magnetic field lines.  Moreover, a few exact
solutions of the stream equation were obtained not only for a zero
longitudinal electric current (Michel 1973a; Mestel \& Wang 1979),
but also for a monopole magnetic field of a
neutron star with a nonzero electric current (Michel 1973b). 
Another solution for
a nonzero electric current was also discussed
(Beskin, Gurevich \& Istomin 1983; Fitzpatrick \& Mestel 1988ab;
Lyubarskii 1990; Sulkanen \& Lovelace 1990). As to the case of inclined
rotator, the solution was obtained for the magnetosphere without a
longitudinal electric current (Henriksen \& Norton 1975;
Beskin et al 1983; see also Mestel \& Wang 1982).
Further on, a more general approach including particle mass was
discussed as well (Ardavan 1979; Bogovalov 1990, 1991; Mestel \& Shibata
1994).

However, the full analysis of the stream equation is far from being completed
even for the force-free axisymmetric magnetosphere.  The point is that the
stream equation describing the pulsar magnetosphere in the presence of a
longitudinal electric current becomes nonlinear. As a result, the
comprehensive analysis of this equation meets certain problems
(Michel 1991; Beskin, Gurevich \& Istomin 1993).  Nevertheless, in some
cases the stream equation can be analyzed.

In this work we consider the simplest approach in which the returning current
flows in a narrow layer along the separatrix, which is the boundary between
open and closed field line regions, and the current
density in the open-line region is constant. In this case, the stream
equation describing the magnetic field structure remains linear not only in a
closed, but in an open magnetosphere as well. As a result, it is possible to
write the general solutions describing the magnetic
field structure in both regions, the main problem lying in matching these
solutions along the separatrix. We shall show that the stability conditions on
the separatrix allow us to obtain a unique solution of the problem.
Moreover, the hypothesis of the compatibility condition (Beskin et al 1983)
connecting the potential drop in the pair creation region and the
longitudinal electric current flowing in the magnetosphere will be confirmed.

\section{Basic equations}

\vspace{0.5cm}

Let us consider an axisymmetric stationary magnetosphere filled with plasma.
This means a zero longitudinal electric field in the
magnetosphere
\beq
({\bmath E}{\bmath B})=0.
\nonumber
\eeq
In this case, the magnetic and electric fields can be represented as
\be
{\bmath {B}}=\frac{\nabla f\times
{\bmath e_\varphi}}{\varpi}+ \frac{\Omega g}{c\varpi}{\bmath e_\varphi},
\label{N1}
\ee
\be
{\bmath {E}}=-\frac{\Omega_{F}}{c}\nabla f,
\label{N2}
\ee
where $c$ is the velocity of
light, and $\Omega$ is the angular velocity of the star. Here $f(\varpi,Z)$ is
the flux function, which is constant along magnetic field lines,
and $g(\varpi,Z)$
determines the whole electric current $I$ flowing inside the tube
$f(\varpi,Z)=const$:
\beq
g=\frac{2I}{\Omega}.
\nonumber
\eeq
Finally, the value
$\Omega_{F}(\varpi, Z)$ is the angular velocity of the plasma. Indeed,
determining the drift velocity from equations~(\ref{N1}) and~(\ref{N2}) we have
\beq
{\bmath
{U}}_{dr}=c\frac{[{\bmath E}\times {\bmath B}]}{B^2}= \Omega_{F}{\bmath e_z}\times
{\bmath r}+i_\parallel{\bmath B},
\nonumber
\eeq
where $i_\parallel$ is a scalar function. Hence, the particle velocity
is a sum of the corotation velocity and the slide velocity along magnetic
field lines.  Using now the Maxwell equation
\beq
[\nabla \times{\bmath E}]=0,
\nonumber
\eeq
which holds
in the stationary case, one can see that the angular velocity $\Omega_F$ must
be constant on the magnetic surfaces $f=const$
\be
\Omega_{F}=\Omega_{F}(f).
\label{N3}
\ee
This is a well-known Ferraro izorotation low (Alfven \& F$\ddot a$lthammar
1963).  It is clear that $\Omega_{F}$ can differ from the angular velocity
$\Omega$ of the neutron star
only on field lines passing through the polar cap region
with a longitudinal electric field, i.e. through the particle creation region.
As a result, in the open magnetosphere the angular velocity $\Omega_F$ must be
smaller than that of the star. On the other hand, we have
$\Omega_{F}=\Omega$ in the region of closed field lines.

Inserting now the electric and magnetic fields~(\ref{N1}) and~(\ref{N2}) into the force-free
condition
\be
{\bmath E}\cdot\nabla{\bmath E}+[\nabla \times{\bmath B}]\times{\bmath B}=0,
\label{N4}
\ee
we obtain from
the $\varphi$ component of equation~(\ref{N4})
\beq
\nabla f\times \nabla g=0.
\nonumber
\eeq
Thus, the electric current $g(\varpi,Z)$ must be constant on the
magnetic surfaces as well
\be
g=g(f).
\label{N5}
\ee
As to the other components of
equation~(\ref{N4}), they just give us the stream equation describing the stable
configuration of the magnetic field lines.  In the dimensionless form
\beq
x=\Omega \varpi/c; \quad z=\Omega Z/c; \quad \omega =
\Omega_{F}/\Omega
\nonumber
\eeq
we have
\be
-\Delta
f\left(1-\omega^{2}x^{2}\right)+\frac{2}{x}\frac{\partial f}{\partial x}-
g\frac{{\rm d}g}{{\rm d}f}+
x^{2}\omega\frac{{\rm d}\omega}{{\rm d}f}\left(\nabla f\right)^{2} =0.
\label{N6}
\ee

This is the general form of the force-free equation describing a stable
magnetic field configuration in the axisymmetric case (Okamoto 1974).
We see that in
equation~(\ref{N6}) the two last terms depending on the longitudinal electric current
$g(f)$ and angular velocity $\Omega_{F}(f)$ can be nonlinear.  On the other
hand, in the closed region (region 1), where longitudinal electric currents
are absent ($g=0$), and the angular velocity is $\Omega_{F}=\Omega$, the equation
becomes linear (Mestel 1973; Michel 1973a)
\be
-\Delta
f^{(1)}(1-x^2)+\frac{2}{x}\frac{\partial f^{(1)}}{\partial x}=0.
\label{N7}
\ee
Moreover, assuming that in the open field region (region 2)
\be
\Omega_{F}=\Omega(1-\beta_0),
\label{N8}
\ee
\be
g(f)=i_{0}f,
\label{N9}
\ee
where $i_0$ and
$\beta_0$ are constants, we obtain the linear equation
\be
-\Delta
f^{(2)}\left[1-x^{2}(1-\beta_{0})^{2}\right]+\frac{2}{x} \frac{\partial
f^{(2)}}{\partial x}-i_{0}^{2}f^{(2)}=0
\label{N10}
\ee
for the open region as well.
Physically, the parameter
$\beta_{0}$ corresponds to the potential drop $\Psi$ in the double layer near
the star surface, i.e. in the particle creation region
\be
\beta_{0}=\Psi/\Psi_{max},
\label{N11}
\ee
where
\be
\Psi_{max}=\left(\frac{\Omega
R}{c}\right)^{2}RB_{0}
\label{N12}
\ee
($R$ is the neutron star radius, $B_{0}$ is the
magnetic field on the star surface) is the maximum potential drop that can be
realized near the star surface inside the open field line region (Ruderman \&
Sutherland 1975).  For $\Psi\approx\Psi_{max}$, $\beta_{0}\approx 1$ we have
$\Omega_{F}\approx 0$, i.e. there is no plasma rotation on the open field
lines. Actually, the value of $\beta_{0}$ can be determined by the pair
creation process in the double layer near the magnetic polar caps (Ruderman
\& Sutherland 1975; Arons 1983; Gurevich \& Istomin 1985).  On the other
hand, we have
\be
i_{0}=\frac{j}{j_{GJ}},
\label{N13}
\ee
where
\beq
j_{GJ}=\frac{\Omega B}{2\pi}
\nonumber
\eeq
is the Goldreich-Julian current density. The condition~(\ref{N9}) just means that
the current density in the open field region is constant.
Thus, with this approach
we have two parameters $i_0$ and $\beta_0$, the latter being
actually determined from a certain pair creation mechanism near the star surface.
The question arises whether it is possible to determine the
electric current $i_0$ as well.

We note, that the stream equation~(\ref{N6}) is valid only
inside the light surface when $|{\bmath E}| < |{\bmath B}|$. Otherwise, one can
not neglect particle inertia in the force-free equation~(\ref{N4}). That is why,
for example, the region of the closed-line magnetosphere can not spread
beyond the light cylinder $x=1$. The closed-line region is limited by
a zero point of the magnetic field $x_*$, which is the point of
intersection of the
separatrix with the equatorial plane. Therefore, we must impose the condition
$x_*\le 1$ to construct closed-line magnetosphere. As for open-line
magnetosphere, the light surface in this region lies beyond the cylinder
$x=1/(1-\beta_0)$ because of the slowing-down of the magnetosphere rotation.

Both equations~(\ref{N7}) and~(\ref{N10}) contain no coordinate $z$.
As a result, the general
solutions can be represented as (Mestel \& Wang 1979; Beskin et al 1983;
Mestel \& Pryce 1992)
\be
f(x,z)=\int_{0}^{\infty}\varphi(\lambda)R(x,\lambda)\cos\lambda zd\lambda.
\label{N14}
\ee
As to the radial function $R(x,\lambda)$, it satisfy an ordinary
second-order differential equation (see below).

In the open region it is convenient to introduce new variables
\beq
x_{1}=x(1-\beta_{0}); \quad z_{1}=z(1-\beta_{0}).
\nonumber
\eeq
One can check that in
these variables equation~(\ref{N10}) only depends on one parameter
\beq
\alpha_{1}=\frac{i_{0}^{2}}{(1-\beta_{0})^{2}}.
\nonumber
\eeq
Moreover, as was shown by
Beskin et al (1983, 1993), for an open field line region we
must omit all harmonics between $\lambda=0$ and $\lambda=\alpha_{1}^{1/2}$ in
the integration~(\ref{N14}). Otherwise, the magnetic field along the rotation axis
will have sinusoidal oscillations, i.e.  magnetic islands. As a result, the
general solution in an open field region can be rewritten in the form
\be
f^{(2)}(x,z)=\int_{0}^{\infty}\varphi_{2}(\lambda)R_{2}(x_{1},\lambda)
\cos\left[(\lambda^{2}+\alpha_{1})^{\frac{1}{2}}z_{1}\right]d\lambda,
\label{N15}
\ee
where now the radial function $R_{2}(x_{1},\lambda)$ is the solution of the
ordinary differential equation
\beq
\frac{{\rm d}^{2}R_{2}(x_{1},\lambda)}{{\rm d}x_{1}^{2}}-
\frac{1+x_{1}^{2}}{x_{1}(1-x_{1}^{2})}\frac{
{\rm d}R_{2}(x_{1},\lambda)}{{\rm d}x_{1}}+ \nonumber \\
\left(\alpha_{1}\frac{x_{1}^{2}}{1-x_{1}^{2}}
-\lambda^{2}\right)R_{2}(x_{1},\lambda)=0.
\label{N16}
\eeq

\section{Boundary conditions}

\vspace{0.5cm}

Let us now consider the boundary conditions for equations~(\ref{N7}) and~(\ref{N10}).  First
of all, at small distances the solutions of both equations must be matched to
the magnetic field of a neutron star.  More exactly, the normal component of
the magnetic field $B_n$ must be continuous on the star surface (Bogovalov
1991). Hence
\be
f(x,z)=f_{star}(R,\theta); \quad x=R\cos\theta;
~z=R\sin\theta.
\label{N17}
\ee
For example, for a dipole magnetic field we have
\be
f(x,z)=m\frac{x^{2}}{(x^{2}+z^{2})^{3/2}},
\label{N18}
\ee
where $m$ is the magnetic dipole of a neutron star, so $f_{star}(R,\theta)=
\left . m\sin^2\theta\right / R$.

It is also necessary to impose the regularity conditions on the singular
surfaces of equations~(\ref{N10}) and~(\ref{N7})
\be
\frac{2}{x}\frac{\partial
f^{(2)}}{\partial x}-i_{0}^{2}f^{(2)}=0; \quad x=\frac{1}{1-\beta_{0}},
\label{N19}
\ee
\be
\frac{\partial f^{(1)}}{\partial x}=0; \quad x=1.
\label{N20}
\ee
On the other hand,
if the zero point is located inside the light cylinder, it is not necessary
to impose the condition~(\ref{N20}) for the closed-line region.

Next we will impose an obvious symmetry condition for the magnetic field in
the closed field line region
\be
\frac{\partial f^{(1)}}{\partial z}=0; \quad z=0,
\label{N21}
\ee
and the condition of zero magnetic flux on the z-axis
for the open field line region
\be
f^{(2)}=0; \quad x=0,
\label{N22}
\ee
The last two conditions are automatically fulfilled if we
represent the solution by~(\ref{N15}) and $\varphi_2(\lambda)$ does not have any
singularities in the upper half-plane of the complex $\lambda$.

Finally, we must match the solutions in the open and closed regions. First of all,
it means that the boundary of the closed region
$f^{(1)}(x,z)=f_{*}$ must coincide with the position of the open region
field line $f^{(2)}(x,z)=f_{*}$, that is,
\be
f^{(1)}[x,z_{*}(x)]=f^{(2)}[x,z_{*}(x)],
\label{N23}
\ee
where $z=z_{*}(x)$ is
the position of the separatrix $f=f_{*}$.  One can see that our approach at
this point differs from that of Lovelace \& Sulkanen (1990) who proposed the
gap between open and closed regions.  Moreover, as pointed out by Okamoto
(1974) and Lyubarskii (1990), for stability of the separatrix separating open
and close regions it is necessary that the value of $L={\bmath B}^{2}-{\bmath E}^{2}$
be continuous through this separatrix
\beq
L^{(1)}[x,z_{*}(x)]=L^{(2)}[x,z_{*}(x)].
\nonumber
\eeq
Using definitions (1), (2) and relations (8), (11), and (13), we obtain
\beq
x^2 L=|\nabla f^{(1)}|^2 [1-x^2]= \nonumber \\
|\nabla f^{(2)}|^2 [1-x^2(1-\beta_0)^2]+i_0^2 f_*^2 .
\label{N24}
\eeq
This condition results from the integration of the full stream
equation~(\ref{N4}) over the narrow boundary $f=f_{*}$. This is the consequence
of nonlinearity of the general stream equation~(\ref{N6}). As a result, to construct
the solution it is necessary to specify two parameters $i_0$, $\beta_0$,
and the condition~(\ref{N17}) on the star surface.
All the other characteristics, in particular, the position and the structure
of the zero point can be determined from the above boundary conditions.

At the very end, it is necessary to stress that
formally, to fulfil the condition~(\ref{N24}) at the zero
point $x_*$, we can expect $x_*<1$,
${\left.\left ({\partial f^{(1)}}/{\partial x}\right )\right|}_{x=x_*}\ne 0$,
and
${\left.\left ({\partial^2f^{(2)}}/{\partial^2 z}\right )\right|}_{x=x_*}=0$.
In other words, here $|\nabla f^{(1)}|^2 [1-x_*^2]=i_0^2 f_*^2$
( Lyubarskii 1990). Therefore the angle $2\vartheta$ between intersecting
separatrices will equal $2\vartheta=\pi$. However,
there is always a narrow transition layer $\delta f$ along the separatrix.
As a result, the magnetic structure in this region remains X-point as for
a magnetosphere with $i_0=0$ and $\beta_0=0$. The only difference is that, as
one can easily see from~(\ref{N7}), the value ${\left .\triangle f\right |}_{x=x_*}$
is equal to zero if $x_*<1$, so $2\vartheta=\pi/2$.
Indeed, supposing that the returning current density is constant in this layer,
we can write $g(f)$, $\omega(f)$ and $L$ for $f_*-\delta f\le f\le f_*$
as
\beq
g=i_0 (f_*-\delta f)(f_*-f)/{\delta f},
\nonumber
\eeq
\beq
\omega=1-\beta_0(f_*-f)/{\delta f},
\nonumber
\eeq
\beq
x^2L=|\nabla f|^2 \left[1-x^2\left (1-\beta_0
\frac{f_*-f}{\delta f}\right )^2\right]+ \nonumber \\
\frac{i_0^2 (f_*-\delta f)^2}{(\delta
f)^2}(f_*-f)^2.
\nonumber
\eeq
Hence, $L$ is continuous through the transition layer
and the separatrix. On the other hand, if $\delta f$ tends to zero, as
is considered in this work, then
we have to take into account that the condition~(\ref{N24}) is not valid near to
zero point. But this is natural, because the stream equation becomes
essentially two-dimensional near the zero point.

\section{Formulation of the problem and numerical results}

\vspace{0.5cm}

In their previous paper Beskin et al (1983) assumed that despite the
presence of an electric current $i_0\ne 0$ 
in the region of open field lines
the structure of closed
magnetosphere remains the same as for $i_0=0$, $\beta_0=0$. In particular,
the zero point is proposed to lie on the light cylinder $x=1$.
It was shown that for these assumptions
the values $i_0$ and $\beta_0$ must be bounded by the compatibility condition
\be
\beta_0=\beta_{*}(i_0),
\label{N25}
\ee
where
\be
\beta_{*}(i_0)=1-\left(1-\frac{i_0^2}{i_{max}^2}\right)^{1/2}
\label{N26}
\ee
and
$i_{max}\approx 1.58$. 
It means that separatrix field lines $f=f_{\ast}$ in open and closed regions 
coinside if the relation (26) is carried out only.
The dependence~(\ref{N26}) is shown in Fig.~1 by the dashed line.
The relation~(\ref{N26}) can be derived directly from the stream equation~(\ref{N10}).
Indeed, assuming that the field line $f=f_*$ passes near the zero point
$x_*=1$ (where $\partial f^{(2)}/\partial x=0$), we immediately obtain the
condition~(\ref{N26}) with
$i_{max}^2=\left .({\left .\triangle f\right |}_{x=x_*})\right / f_*$.
The goal of the present paper is actually to check the compatibility
condition~(\ref{N25}) in a more self-consistent approach, i.e. to include the
disturbance of the closed region and the stability condition~(\ref{N24}).
\begin{figure}
\vspace{6cm}
\includegraphics{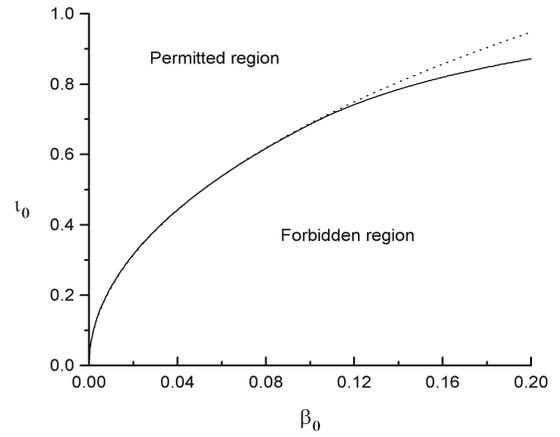}
\caption{
Forbidden and permitted regions in the $i_0 - \beta_0$ diagram.
The solid curve corresponds to the compatibility condition $i_0(\beta_0)$,
while the dashed curve represents the dependence (26).
}
\end{figure}

As has already been shown, equations~(\ref{N7}),~(\ref{N10}) and the boundary 
conditions~(\ref{N17}),~(\ref{N19})--(\ref{N24})
give a solution of the problem if it exists. On the other hand, we do not
know beforehand the position of the separatrix. It can be determined from the
solution. In other words, the function $\varphi_2 (\lambda)$ in~(\ref{N15}), which
determines, in particular, the position of the separatrix and the zero point,
is unknown. However,
the magnetic field has a dipole asymptotic behavior when $x$, $z$ tend to zero. For
the representation~(\ref{N15}) and the star magnetic dipole $m=1$, this assumption
corresponds to
\be
\lim_{x\rightarrow 0}R_2(x_1,\lambda)= \frac{2}{\pi}\lambda xK_1(\lambda x),
\label{N27}
\ee
because we have for a
dipole field~(\ref{N18})
\beq
\frac{x^{2}}{(x^{2}+z^{2})^{3/2}}=\frac{2}{\pi}\int_{0}^{\infty}\lambda x
K_{1}(\lambda x)\cos\lambda zd\lambda.
\nonumber
\eeq
Hence,
\be
\lim_{\lambda\rightarrow \infty}\varphi_2(\lambda)=1.
\label{N28}
\ee
On the other hand, from~(\ref{N19}),~(\ref{N15}) we have for the function $R_2(x_1,\lambda)$
\be
\frac{2}{x_1}\frac{\partial R_2}{\partial x_1}-\alpha_1 R_2=0; \quad x_1=1.
\label{N29}
\ee
As a result, for given parameters $i_0$ and $\beta_0$ the function
$\varphi_2(\lambda)$ completely determines the open-line magnetosphere. Then
using equation~(\ref{N7}) and the boundary condition~(\ref{N23}),~(\ref{N21}) we can determine the
closed-line magnetosphere as well.

In the present work, for any pair of parameters from the $i_0$--$\beta_0$ plane we
are looking for the best function $\varphi_2(\lambda)$ to satisfy the
condition~(\ref{N24}). In other words, to obtain the continuity of 
$L={\bmath B}^{2}-{\bmath E}^{2}$ through the separatrix, for any $i_0$, $\beta_0$ we minimize the
functional
\be
S[\varphi_2(\lambda)]=\max_{0\le x\le x_*-\delta x}
		     \left|L^{(2)}[x,z_*(x)]-L^{(1)}[x,z_*(x)]\right|,
\label{N30}
\ee
where $\delta x$ takes into account that the
condition~(\ref{N24}) is not valid in the vicinity of the zero point.
In more detail, we expand the
function $\varphi_2(\lambda)$ in a finite functional series
\beq
\varphi_2(\lambda)=1+\sum_{n=1}^N C_n \phi_n(\lambda)
\nonumber
\eeq
to reduce the
problem of functional minimization to the problem of minimization of
function with many variables, which are coefficients of the expansion.
In our work we use $N=2$ and
\beq
\phi_1(\lambda)=
\left \{
\begin{array}{ll}
1, & \lambda\in [0.2, 0.6] \nonumber\\
0, & \lambda\notin [0.2, 0.6] \nonumber
\end{array}
\right. ,
\nonumber
\eeq
\beq
\phi_2(\lambda)=\left \{
\begin{array}{ll}
1, & \lambda\in [1.5, 3.0] \nonumber\\
0, & \lambda\notin [1.5, 3.0] \nonumber
\end{array}
\right . .
\nonumber
\eeq
So, the function $\varphi_2(\lambda)$ is different from one on two
intervals $\lambda\in [0.2, 0.6]$ and $\lambda\in [1.5, 3.0]$. The result
does not sufficiently depend on what intervals to choose. The function
$\phi_1(\lambda)$, that is nonzero for smaller $\lambda$--interval, and
the function $\phi_2(\lambda)$, that is nonzero for greater
$\lambda$--interval, allow us to minimize the functional~(\ref{N30}) for $x\sim 1$
and $x\ll 1$ respectively.

As has already been said, the zero point must lie inside the light
cylinder $x_*\le 1$. On the contrary, the open region can not be matched
to the closed one because the closed-line magnetic field may not spread
beyond the light cylinder $x=1$. As a result, a self-consistent solution of the
problem can be obtained not for all the parameters $i_0$ $\beta_0$.
This is illustrated in the $i_0$--$\beta_0$ diagram in Fig.~1. The solid
line is the boundary of the forbidden region of the parameters $i_0$, $\beta_0$.
For the forbidden region, the zero point of the magnetic field moves outside
the light cylinder.

As for the permitted region, numerical calculations show that for a
given $\beta_0$ we get the best
conformity to the condition~(\ref{N24}) if we take $i_0$ that lies just on the boundary
between the forbidden and the permitted regions. Simultaneously, the zero
point of the magnetic field just lies on the light cylinder $x_*=1$.
Moreover, the minimum of the electromagnetic field energy
\beq
{\it E_{em}}=\int\limits_{0}^{x_*}\int\limits_{0}^{z_*(x_*)} x^{-1}(1+x^2)
|\nabla f^{(1)}|^2 dz dx+
\nonumber \\
\int\limits_{0}^{x_*}\int\limits_{z_*(x_*)}^{\infty}
x^{-1}\left [ {\left |\nabla f^{(2)}\right |}^2 \left (1+x^2(1-\beta_0)^2\right ) +
i_0^2 {\left ( f^{(2)}\right ) }^2 \right ] dz dx
\nonumber
\eeq
corresponds to the boundary curve as well.
To illustrate the above statements we show the dependence of the zero point
$x_*$, the value of the functional $S[\varphi_2(\lambda)]$ (30) corresponded to the
drop of $L={\bmath B}^2-{\bmath E}^2$ and
the electromagnetic field energy ${\it E_{em}}$
from $i_0$ for the $\beta_0=0.05$ in Figs.~2a,~2b~and~2c respectively.
As is seen from Fig.~2b, we can not find the exact function $\varphi_2(\lambda)$
to obtain zero drop of $L$ because we would have to expand
$\varphi_2(\lambda)$ in an infinite functional series in this case.
\begin{figure}
\vspace{18cm}
\includegraphics{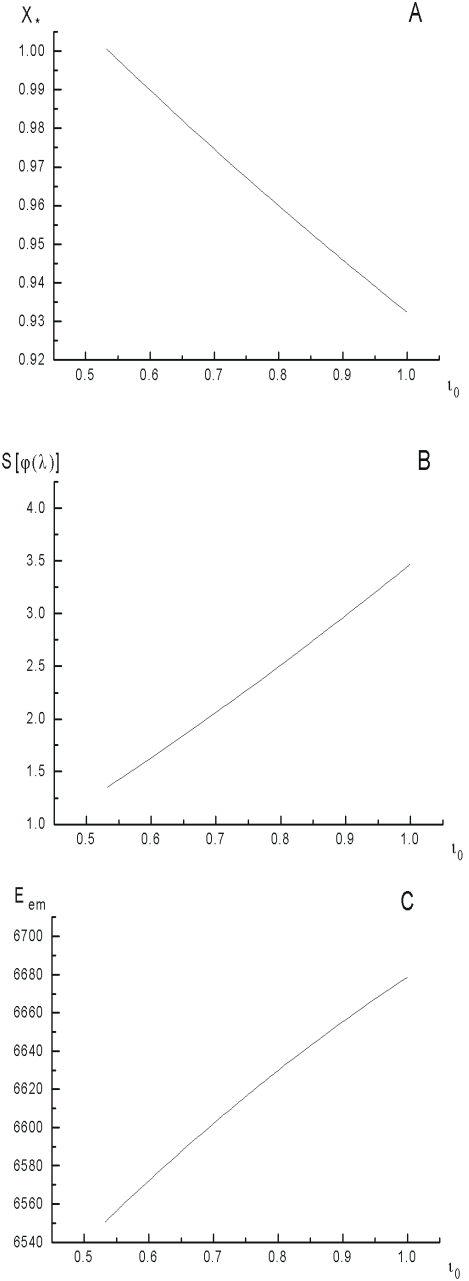}
\caption{
The dependence of the zero point $x_*$ (A), the value of the 
functional $S[\varphi_2(\lambda)]$ (30) corresponded to the
drop of
$L={\bmath B}^2-{\bmath E}^2$ (B) and the electromagnetic field energy
${\it E_{em}}$ (C) from $i_0$ for the $\beta_0=0.05$. One can see that
for $i_0=0.52$, which corresponds to the compatibility condition, we have
$x_*=1$ and the minimum of the values $S[\varphi_2(\lambda)]$, $E_{em}$.
}
\end{figure}

Thus, we can conclude that the compatibility condition $\beta_0=\beta_{*}(i_0)$
does actually exist. It is represented by the boundary solid line in Fig.~1.
One can see that the difference between the dashed and solid lines is very
slight for small values of $i_0$, $\beta_0$, and hence we conform the
dependence~(\ref{N26}) for a small current density and a potential drop.

The structure of the magnetosphere  for $\beta_0=0.05$ and
$i_0=0.52$, which corresponds to the compatibility condition,
is shown in Fig.~3.
We see that the zero point of the magnetic field lies just on the light
cylinder $x=1$
and the angle $2\vartheta$ between the intersecting separatrices corresponds to
$\tan^2 \vartheta=1/2$. Therefore, we confirm the hypothesis by Beskin et al
(1983) that the shape of the closed region in the presence of all electric
current and a potential drop does not differ greatly from that for the zero
current. On the other hand, in Fig.~4 we show an example, where
the zero point lies inside the light cylinder ($x_*< 1$). This can be
realized for the parameters $i_0$ and $\beta_0$ far from the compatibility
condition.
\begin{figure}
\vspace{6cm}
\includegraphics{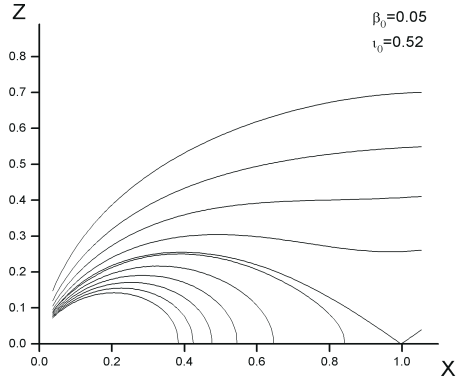}
\caption{
The structure of the axisymmetric magnetosphere of a neutron star for
the potential drop and the current density which are connected by the
compatibility condition. The zero point is located on the light cylinder
$x=1$ and the angle $2\vartheta$ between the intersecting separatrices
corresponds to $\tan^2 \vartheta=1/2$.
}
\end{figure}
\begin{figure}
\vspace{6cm}
\includegraphics{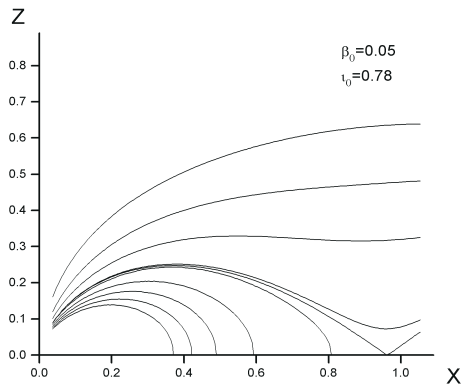}
\caption{
The structure of the axisymmetric magnetosphere of a neutron star for
the potential drop and the current density which are far away from the
compatibility condition. The zero point is located inside the light cylinder
$x=1$.
}
\end{figure}

\section{Discussion}

\vspace{0.5cm}

Thus, we have shown that the compatibility condition between current and
potential drop remains true even if we include the boundary 
condition~(\ref{N24}) into consideration. 
On the other hand, it was shown by Beskin et al (1983) that
the relation~(\ref{N25}) remains, in general, the same for an inclined rotator
as well. 
For example, the compatibility relation has now the form
\be
\beta_0(i_0)
=\beta_{max}(\chi)-\left(1-\frac{i_0^2}{i_{max}^2(\chi)}\right)^{1/2},
\ee
where 
$\beta_{max}$ and $i_{max}$ depend now on the inclination angle~$\chi$.

As a result, the longitudinal current flowing in the neutron star
magnetosphere is not a free parameter and determined by the
particle creation mechanism in the polar regions. On the other hand,
it is the longitudinal current that causes deceleration of pulsar
rotation. For example, the total energy losses in the model proposed by
Beskin et al (1983) is
\be
{\it I_r}\Omega{\dot \Omega}=W_{tot}=\frac{f_*^2(\chi)}{8}\frac{B_0^2 \Omega^4
R^6}{c^3} i_0(\beta_0) \cos\chi,
\label{N31}
\ee
where $\chi$ is the angle of pulsar inclination, $\it I_r$ is a moment of
inertia of the pulsar and $1.59<f_*(\chi)<1.95$.

The full analysis of observational data goes beyond the scope of this paper.
However, we briefly recall the main predictions which result from the
existence of the Ohm's Law~(\ref{N25}).

1. As we understand the nature of pulsars, their radio emission results
from a secondary plasma which is produced by a longitudinal electric field
near the pulsar surface. Therefore, the condition $\Psi(P,B)=\Psi_{max}$
just determines the maximum period of radio pulsars. For example, in the model
with large enough work function
(Ruderman \& Sutherlend 1975; Gurevich \& Istomin 1985) we have
$P_{max}\approx 1s$, which is in agreement with observations.
Using the compatibility condition $i_0=i_0(\beta_0)$ and~(\ref{N11}),
(\ref{N13}),~(\ref{N31}), we can rewrite the inequality $P < P_{max}$ in the form $Q < 1$,
where
\beq
Q=2\left(\frac{P}{1c}\right)^{11/10}
\left(\frac{\dot P}{10^{-15}}\right)^{-4/10}.
\nonumber
\eeq
The parameter $Q\approx i_0$, which can be determined from observations,
is very convenient for expressing the basic pulsar characteristics
(Beskin et al 1984; Taylor \& Stinebring 1986; Rankin 1990).
For example,
in the hollow cone model the ratio of the inner radius $r_{in}$ and the
height $H$ of plasma generation region to the polar cap radius $R_0$ of a
pulsar is
\beq
\frac{r_{in}}{R_0}\sim Q^{7/9};
\qquad \frac{H}{R_0}\sim Q.
\nonumber
\eeq
Thus, we can conclude that pulsars with $Q\approx 1$ have a thin cone of
emission and hence they have a two-peak radio emission profile. It is these
pulsars that have emission nonregularity such as nulling, mode switching, etc.
On the contrary, pulsars with $Q \ll 1$ have a stable one-peak profile
emission. Such situation is just corresponds to observations
(Taylor \& Stinebring 1986; Beskin et al 1993).

2. One of the main radio pulsar parameters that characterizes
pulsar rotation deceleration is a braking index
$n_{br}=\left .\Omega{\ddot\Omega}\right /{\dot\Omega^2}$
which is directly available from observations (Lyne \& Graham-Smith 1990).
Unfortunately, the braking index is only known for a few radio pulsars.
In particular, $n_{br}=2.24$ for PSR $0540-693$ and $n_{br}=2.84$ for
PSR $1509-58$. In the case of vacuum magnetodipole
energy losses we have $n_{br}=3+2\tan^{-2}\chi$,
which can not explain the observations.
On the other hand, for the current losses~(\ref{N31}) and the compatibility relation
$i_0(\beta_0)$~(\ref{N26}) we have $n_{br}=1.93+1.5\tan^2\chi$, which is in good
agreement with observational data.

3. In the Ruderman-Sutherlend model (1975) the longitudinal current is equal
to the Goldreich-Julian one. Hence, the energy of particles hitting against the
pulsar polar cap is high enough to cause an intensive X-ray emission. This
contradicts observational data. Nevertheless, the X-ray emission is only
observed to fast-period pulsars which have a considerably smaller current.
Then, in accordance with the model discussed and the compatibility
condition, the energy of pulsar cap heating is $W_{X}\approx Q^{2}W_{tot}$
(Beskin et al 1993),
which is significantly less then the generally accepted estimates because
for these pulsars $Q^2\approx i_0^2=\left (j/j_{GJ}\right )^2 \ll 1$. It is
necessary to stress that in the Arons model (1993) with a small work function
this difficulty is absent.

4. If a pulsar energy lost is caused by longitudinal current, then the
braking angular momentum $\bmath K$ is opposite to the magnetic dipole of a
neutron star. As a result, the value $\Omega\sin\chi$ remains constant during
radio-pulsar evolution. Consequently the angle $\chi$ between dipole axis
and rotation axis of the pulsar is increasing during pulsar life, while
for magnetodipole energy lost it is decreasing.

Thus, as we see, the main predictions of the theory with a large work function
and the compatibility condition $\beta_0=\beta_*(i_0)$ at least do not
contradict observational data. Moreover, to confirm the theory, there have
recently appeared some indirect observational results such as the absence of
evolution of radio-pulsar magnetic field and the statistical conclusion that
the initial pulsar periods are about $0.1-1$ s rather than $1-10$ ms as
for Crab and Vela young pulsars. Of course, the direct measurement of
the evolution of pulsar inclination angle $\chi$
%, which tends to $\pi/2$ for current losses, 
will be the key-experiment. Unfortunately, such an experiment
is impossible to carry out at the present time.

\begin{center}
{\bf Acknowledgments}
\end{center}
The authors thank Ya.N.Istomin, I.Okamoto, N.Shibazaki, and S.Shibata for
fruitful discussions. This work was supported by Center for International
Studies, Rikkyo University, Tokyo and INTAS Grant 94-3097. LM also thanks
the International Soros Student Educational Program for financial support.

\end{document}